\documentclass[12pt, letterpaper]{article}

\usepackage{amsmath}
\usepackage{amsfonts}
\usepackage{amssymb}

\DeclareMathOperator{\sech}{sech}
\DeclareMathOperator{\csch}{csch}

\begin{document}

\begin{flushright} {\footnotesize YITP-18-77, IPMU18-0122}  \end{flushright}
\vspace{5mm}
\vspace{0.5cm}
\begin{center}

\def\thefootnote{\fnsymbol{footnote}}

{\Large {\bf Gauge Field Mimetic Cosmology}}
\\[1cm]

{ Mohammad Ali Gorji$^{1}\footnote{gorji@ipm.ir}$, Shinji Mukohyama$^{2,3}\footnote{shinji.mukohyama@yukawa.kyoto-u.ac.jp}$, \\
 Hassan Firouzjahi$^{1}\footnote{firouz@ipm.ir}$, Seyed Ali Hosseini 
 Mansoori$^{4}\footnote{shosseini@shahroodut.ac.ir}$ }
\\[.7cm]

{\small \textit{$^1$School of Astronomy, Institute for Research in Fundamental Sciences (IPM) \\ P.~O.~Box 19395-5531, Tehran, Iran
}}\\

{\small \textit{$^2$Center for Gravitational Physics, Yukawa Institute for Theoretical
Physics, Kyoto University, 606-8502, Kyoto, Japan}}\\

{\small \textit{$^3$Kavli Institute for the Physics and Mathematics of the Universe (WPI),
The University of Tokyo Institutes for Advanced Study, The University of Tokyo,
Kashiwa, Chiba 277-8583, Japan }}\\

{\small \textit{$^4$
Faculty of Physics, Shahrood University of Technology,\\ P.O.Box 3619995161 Shahrood, Iran }}

\end{center}

\vspace{.8cm}

\hrule \vspace{0.3cm}

\begin{abstract} 
We extend the mimetic cosmology to models containing  gauge invariant $p$-forms. The $0$-form case reproduces the well-known results of the mimetic dark matter, the $1$-form corresponds to the gauge field mimetic model while the $2$-form model is the Hodge dual of the $0$-form model in $4$ spacetime dimensions. We  study the cosmological applications of the new gauge field mimetic model and show that it generates an energy density component which mimics the  roles of spatial curvature. In the presence of the Maxwell term, the model also supports the flat, open and closed de Sitter-like cosmological backgrounds while the spatial geometry is flat for all three cases. We perform the cosmological perturbations analysis and show that the model is stable in the case of open de Sitter-like solution while it suffers from ghost instabilities in the case of the closed de Sitter-like solution. 
\end{abstract}
\vspace{0.5cm} 
\hrule
\def\thefootnote{\arabic{footnote}}
\setcounter{footnote}{0}

\newpage
\section{Introduction}

Recently, the mimetic gravity has been suggested as a modification of general relativity such that 
one isolates the conformal degree of freedom of the metric by means of a scalar field 
\cite{Mimetic-2013}. The setup can be equivalently realized from degenerate conformal/disformal 
transformation. The number of degrees of freedom then increases such that the longitudinal 
mode of gravity turns out to be dynamical \cite{Deruelle:2014zza,Firouzjahi:2018xob}. Evidently, this new degree 
of freedom can play the roles of dark matter in cosmological setup \cite{Mirzagholi:2014ifa}. 
Apart from these interesting features of the mimetic dark matter model (see Refs. 
\cite{mimetic-recent} for the recent works), the scalar mode is
plagued with instabilities \cite{Ramazanov:2016xhp, Ijjas:2016pad, Firouzjahi:2017txv} and without taking into account some coupling between the second derivative of the scalar field and curvatures, one cannot find healthy
perturbations \cite{Hirano:2017zox,Gorji:2017cai,Langlois:mimetic}. The model also suffers from caustics that are 
formed virtually everywhere in the universe \cite{caustics,Barvinsky:2013mea}. 

Here, we would like to seek for the possibility of isolating the conformal degree of freedom of 
gravity by means of gauge fields rather than a scalar field. Some attempts have been made in 
this direction. In Ref. \cite{Barvinsky:2013mea}, it is argued that the usual scalar-tensor 
mimetic gravity cannot admit rotating dark matter while the vector-tensor counterpart can 
simulate the rotating flows of dark matter. It is also shown that the vector-tensor extension 
of the mimetic gravity would be free of ghost instabilities \cite{Chaichian:2014qba}.
In order to construct mimetic vector-tensor model via gauge fields, we briefly review the basic 
idea of the mimetic scalar-tensor gravity from the conformal/disformal transformation point of view. 

Consider the conformal transformation $g_{\mu\nu}=\, A(\phi,K) {\tilde g}_{\mu\nu}$ between the 
physical metric $g_{\mu\nu}$ and an auxiliary metric $\tilde{g}_{\mu\nu}$ in which the 
coefficient $A>0$ being the function of scalar field $\phi$ and its canonical kinetic term 
$K = -\tilde{g}^{\alpha\beta} \partial_{\alpha}\phi\partial_{\beta}\phi$\footnote{We work in the mostly positive metric signature $(-,+,+,+)$.}. Then the nontrivial singular limit of the transformation uniquely fixes the 
functional form of the conformal coefficient as $A=K$ 
\cite{Deruelle:2014zza,Zumalacarregui:2013pma} (up to an overall factor independent of $K$) which implies the following transformation
\begin{eqnarray}\label{mimetic-trans}
g_{\mu\nu}=( -\tilde{g}^{\alpha\beta} \partial_{\alpha}\phi\partial_{\beta}\phi )\, 
{\tilde g}_{\mu\nu} \, .
\end{eqnarray}

It is easy to see that the physical metric $g_{\mu\nu}$ is invariant under the conformal 
transformation of the auxiliary metric ${\tilde g}_{\mu\nu}$ and also satisfy the following 
constraint \cite{Golovnev:2013jxa}

\begin{eqnarray}\label{mimetic-scalar-cons.}
g^{\alpha\beta} \partial_{\alpha}\phi\partial_{\beta}\phi = -1 \, .
\end{eqnarray}
The transformation (\ref{mimetic-trans}) is a particular (conformal) case of more general 
transformations known as disformal transformations \cite{Bekenstein:1992pj}. The number of 
degrees of freedom does not change under non-degenerate disformal transformation 
\cite{Domenech:2015tca}. However, the mimetic transformation (\ref{mimetic-trans}) is degenerate and 
the number of degrees of freedom increases such that the scalar field $\phi$ is present even in 
the absence of usual matter and makes the longitudinal mode of gravity dynamical. 

On the other hand, one can replace $\partial_{\mu}\phi$ with a vector field $A_{\mu}$ and construct a mimetic 
vector-tensor gravity \cite{Barvinsky:2013mea} (see also Ref. \cite{Vikman:2017gxs}). These 
models can be equivalently obtained from the following mimetic (singular) vector transformation 
$g_{\mu\nu}=(-\tilde{g}^{\alpha\beta} A_{\alpha} A_{\beta} )\, {\tilde g}_{\mu\nu}$, provided that $A_{\mu}$ (instead of $A^{\mu}$) is transformed to itself. The model then implies constraint $g^{\mu\nu} A_{\mu} A_{\nu} = -1$ which is the well-known
constraint for Einstein-Aether model \cite{Jacobson:2000xp}. This model however clearly breaks 
the local gauge invariance in the corresponding action. In the present paper we are interested in a 
gauge invariant generalization of the mimetic scenario. We therefore consider conformal 
transformation $g_{\mu\nu}=\, A(X) {\tilde g}_{\mu\nu}$ in which $X=-\tilde{g}^{\rho\alpha} 
\tilde{g}^{\sigma\beta} F_{\alpha\beta} F_{\rho\sigma}$ is the standard Maxwell term and $F_{\mu\nu}$ is the strength tensor associated to the gauge field. Following the same 
step as in the case of scalar field, it is straightforward to show that the singular limit of the 
transformation is given by (see the appendix A for the detail)
\begin{eqnarray}\label{mimetic-trans-GF}
g_{\mu\nu}=\left(-\tilde{g}^{\rho\alpha} \tilde{g}^{\sigma\beta} F_{\alpha\beta} F_{\rho\sigma}
\right)^{\frac{1}{2}}\, {\tilde g}_{\mu\nu} \, .
\end{eqnarray} 
Note that the physical 
metric $g_{\mu\nu}$ is invariant under the conformal transformation of the auxiliary metric 
${\tilde g}_{\mu\nu}$ in (\ref{mimetic-trans-GF}). Apart from systematic derivation of 
(\ref{mimetic-trans-GF}) in the appendix A, it is easy to understand why the square root of the 
scalar $F^2$ has appeared in (\ref{mimetic-trans-GF}): the inverse metric enters twice in 
$\tilde{g}^{\rho\alpha} \tilde{g}^{\sigma\beta} F_{\alpha\beta} F_{\rho\sigma}$ while it appears once in 
the case of $\tilde{g}^{\alpha\beta} \partial_{\alpha}\phi\partial_{\beta}\phi$.

The inverse of the metric (\ref{mimetic-trans-GF}) is given by $g^{\mu\nu}=\left(
- \tilde{g}^{\rho\alpha} \tilde{g}^{\sigma\beta} F_{\alpha\beta} F_{\rho\sigma}
\right)^{-\frac{1}{2}}\, {\tilde g}^{\mu\nu}$. Multi-plying it with another inverse metric and then
contracting the result with two strength tensors, it is easy to show that the conformal mimetic 
gauge field transformation (\ref{mimetic-trans-GF}) implies the following constraint
\begin{eqnarray}\label{mimetic-const.0}
g^{\mu\alpha} g^{\nu\beta} F_{\alpha\beta} F_{\mu\nu} = - 1 \,.
\end{eqnarray}
The above constraint is the gauge-invariant vector extension of the scalar mimetic constraint 
(\ref{mimetic-scalar-cons.}).

Note that the scalar mimetic constraint (\ref{mimetic-scalar-cons.}) is invariant under the shift 
symmetry $\phi\rightarrow\phi+c$ where $c$ is a constant while the new proposed mimetic 
constraint (\ref{mimetic-const.0}) is invariant under the gauge symmetry $A_{\mu} \rightarrow 
{A}_{\mu} - \partial_{\mu} \Lambda$, where $\Lambda$ is an arbitrary function. Considering 
$\phi$ to be a $0$-form and $A_{\mu}$ to be the components of a $1$-form, the goal of this work is to 
generalize the results (\ref{mimetic-scalar-cons.}) and (\ref{mimetic-const.0}) to a general 
$p$-form in a unified manner.

The rest of the paper is organized as follows. In Section \ref{pform} we extend the mimetic constraint to the general case of gauge invariant $p$-form. In Section \ref{1-form-sec} we study the cosmological applications of 
the gauge field mimetic model with global $O(3)$ symmetry, corresponding to the case $p=1$. We study both the background dynamics and the cosmological perturbations. The summary of the paper and some discussions are presented in Section \ref{summary}.  The analysis of singular conformal transformation which motivates our mimetic constraints for $p$-form gauge fields are presented in Appendix A while the equivalence of the two models with $p=0$ and $p=2$ is demonstrated in Appendix B. 
 

\section{Gauge-invariant Mimetic p-forms}
\label{pform}
In this section, our aim is to formulate the general mimetic p-form constraint which reduces to the constraints 
(\ref{mimetic-scalar-cons.}) and (\ref{mimetic-const.0}) for the special cases $p=0$ and $p=1$ respectively.

Consider the $p$-form potential ${\cal A}$ on the Lorentzian manifold $({\cal M},g)$ with $g$ being the metric. The associated field strength will be ${\cal F}=d{\cal A}$ where $d$ denotes the exterior derivative. The gauge-invariant Yang-Mills Lagrangian is given by ${\cal F} \wedge \star {\cal F}$. We define $\langle {\cal F},{\cal F}\rangle$ by ${\cal F} \wedge \star {\cal F}=\langle {\cal F},{\cal F}\rangle \, \star 1 /(p+1)!$. For $p=0$ and with the scalar potential $\phi$, we find $\langle {\cal F},{\cal F}\rangle = \partial^\mu\phi \partial_\mu\phi$ which is the standard kinetic term for the scalar field. In the same manner and for $p=1$ we find $\langle {\cal F},{\cal F}\rangle = F_{\mu\nu}F^{\mu\nu}$ where $F_{\mu\nu}=\partial_\mu A_\nu-\partial_\nu A_\mu$. The standard mimetic constraint for the scalar field is given by $\partial^\mu\phi \partial_\mu\phi = -1$. Having this in mind we propose mimetic p-form constraint to be 
\begin{eqnarray}\label{mimetic-const-p}
\langle {\cal F},{\cal F} \rangle = \mp 1 \,.
\end{eqnarray} 
For $p=0$ and minus sign, the above constraint reduces to the standard scalar mimetic 
constraint (\ref{mimetic-scalar-cons.}) and for $p=1$ and minus sign, it reduces to the gauge 
field mimetic constraint (\ref{mimetic-const.0}). The plus sign corresponds to other possibilities.

Having the $p$-form mimetic constraint (\ref{mimetic-const-p}) in hand, we can easily define 
the gauge-invariant mimetic p-form model as
\begin{equation}\label{action-pform}
S_p = \frac{1}{2} \int_{\cal M} \left[\star R - \lambda_p 
\big({\cal F} \wedge \star{\cal F} \pm \star 1\big)\right] 
= \frac{1}{2} \int_{\cal M} d^{4}x \sqrt{-g} \left[ R - \lambda_p 
\big(\langle {\cal F},{\cal F} \rangle\pm1\big) \right] \,,
\end{equation}
where $\lambda_p$ are auxiliary fields which enforce the mimetic constraint 
(\ref{mimetic-const-p}). 

Note that apart from the cosmological constant term, we can also add a general function 
$f(\langle {\cal F},{\cal F} \rangle)$ to the above action which is forced to be constant by the 
mimetic constraint (\ref{mimetic-const-p}). In some sense, this term plays the roles of 
cosmological constant. We will explicitly see this fact in the next section.

Let us consider the action (\ref{action-pform}) for various values of $p$. 
For the case $p=0$, the above action gives
\begin{equation}\label{action-p0}
S_0 = \frac{1}{2} \int d^{4}x \sqrt{-g}\Big[\, R -\lambda_0 
\big(\partial^\mu\phi \partial_\mu\phi\pm1\big)\,\Big]\,.
\end{equation}
With the plus sign, the above action is nothing but the action of the mimetic dark matter model
\cite{Golovnev:2013jxa}. In the case of minus sign, the vector $\partial_\mu\phi$ is space-like 
while it is time-like in standard mimetic scenario and plays the roles of the gradient of the 
velocity potential \cite{Mimetic-2013}.

For the case of $p=1$, we have
\begin{equation}\label{action-p1}
S_1 = \frac{1}{2} \int d^{4}x \sqrt{-g}\Big[\, R - \lambda_1 
\Big( F_{\mu\nu} F^{\mu\nu} \pm 1 \Big) \,\Big].
\end{equation}
Since the electric field gives negative contribution to the $F^2$ term and the magnetic field gives
positive contribution, depending on the situation we can consider both the positive and negative 
signs. We will see that considering three $U(1)$ guage-invariant vector fields and assuming a global $O(3)$ symmetry in the field space, the negative sign with purely electric contribution admits cosmological solution. 

For the case of $p=2$ and the potential $2$-form $B=\frac{1}{2!}B_{\mu\nu}dx^{\mu}\wedge 
dx^{\nu}$, we have
\begin{equation}\label{action-p2}
S_2 = \frac{1}{2} \int d^{4}x \sqrt{-g}\Big[\, R - \lambda_2
\Big( J_{\mu\nu\gamma} J^{\mu\nu\gamma} \pm 1 \Big) \,\Big]\,,
\end{equation}
where $J_{\mu\nu\gamma}=\partial_{\mu}B_{\nu\gamma} + \partial_{\nu}B_{\gamma\mu} + \partial_{\gamma}B_{\mu\nu}$ is the component of the field strength $3$-form $J=dB$.

Finally, in the case of $p=3$ the field strength is a $4$-form, known as top form, and is 
proportional to the Levi-civita tensor. The corresponding mimetic constraint then only fixes the magnitude of a non-dynamical field which does not give any special result.

An interesting fact here is that, similar to the free field p-form models \cite{Thorsrud:2017rpl}, 
the cases $p=0$ with the action (\ref{action-p0}) and $p=2$ with the action (\ref{action-p2}) are 
physically equivalent. In appendix B, we have shown that, at the level of equations of motion, 
these two models are related to each other through the Hodge duality. Therefore, the models 
(\ref{action-p0}) and (\ref{action-p2}) with $p=0$ and $p=2$ 
describe the well-known mimetic dark matter scenario which was extensively studied in the literature and we do not consider them any further here. 

What remains is the case (\ref{action-p1}) with $p=1$  which we are interested in this paper. 
This model is gauge-invariant and can be considered as the mimetic extension of the standard Einstein-Maxwell model. However, we are interested in cosmological applications of the model and for this purpose it turns out that the setup can be extended to accommodate a global $O(3)$ symmetry as we shall study in the next section.


\section{1-form model: Cosmological Implications}
\label{1-form-sec}

There is only one gauge field in the mimetic $1$-form model (\ref{action-p1}) which intrinsically carries a privileged direction and therefore inevitably introduces anisotropy\footnote{One interesting case is to consider Bianchi geometries in this setup which is not the purpose of this paper.}. We are interested in cosmological applications of the model (\ref{action-p1}) but the model does not admit the  isotropic FRW background solution. A natural extension of (\ref{action-p1}) that makes the model compatible with homogeneity and isotropy is to consider an orthogonal triplet of gauge fields \cite{Golovnev:2008cf} such that the field space enjoys a global $O(3)$ symmetry 
\cite{Emami:2016ldl}
\begin{equation}\label{action}
S=\int d^{4}x \sqrt{-g}\Big[\, \frac{1}{2} R - \lambda \Big(\sum_{a=1}^{3}
F^{(a)}_{\mu\nu} F^{(a) \mu\nu}+1 \Big) - \frac{1}{4} \, {\mathcal E} \,
\sum_{a=1}^{3} F^{(a)}_{\mu\nu} F^{(a)\mu\nu} \,\Big] \,,
\end{equation}
where
\begin{eqnarray}\label{Fmunu-definition}
F^{(a)}_{\mu\nu}=\partial_{\mu}A^{(a)}_{\nu}
-\partial_{\nu}A^{(a)}_{\mu} \,,
\end{eqnarray}
$\mu,\nu=0,..,3$ are spacetime indices while $a=1,2,3$ denotes field space indices. The auxiliary field $\lambda$ enforces the mimetic constraint
\begin{eqnarray}\label{mimetic-const.}
\sum_{a=1}^{3} g^{\mu\alpha} g^{\nu\beta} F^{(a)}_{\alpha\beta} F^{(a)}_{\mu\nu} = - 1 \,.
\end{eqnarray}

The generalization of the calculations of the appendix A to the case of global $O(3)$ 
symmetry in field space is straightforward such that one can obtain the above constraint 
from the singular limit of the conformal transformation $g_{\mu\nu} = A(X) \tilde{g}_{\mu\nu}$ 
with $X = - \sum_{a=1}^{3} g^{\mu\alpha} g^{\nu\beta} F^{(a)}_{\alpha\beta} F^{(a)}_{\mu\nu}$. 

In the action (\ref{action}) we have also considered the Maxwell-like term (the term containing ${\mathcal E}$) which, as we have already mentioned, would play the roles of cosmological constant through the mimetic constraint 
(\ref{mimetic-const.}). Indeed, at the level of the action, it is obvious that one can replace the Maxwell-like term by a constant term via a redefinition of the Lagrange multiplier $\lambda$. Having a term like a cosmological constant term signals the existence of de Sitter-like solution. We will explicitly show this fact in the rest of this section by means of the Raychuadhuri and the Friedmann equations.

Varying the action (\ref{action}) with respect to $A^{(a)}_{\mu}$, we obtain Maxwell-like equations 
$\nabla_{\mu} \Big(({\mathcal E}+4\lambda) F^{(a)\mu\nu}\Big)=0$. Varying the action with respect to the metric, one obtains the Einstein field equations 
$G^\mu_\nu = T^\mu_\nu$ (in the unit where the reduced Planck mass $M_P$ is set to unity), where $G^\mu_\nu$ is the Einstein tensor and $T^\mu_\nu$ is the effective
energy momentum tensor given by
\begin{equation}\label{EMT}
T^{\mu}_{\nu}=\frac{\mathcal E}{4}\,\delta^{\mu}_{\nu}
+ ({\mathcal E}+4 \lambda) \sum_{a=1}^{3} F^{(a)\mu\alpha}
F^{(a)}_{\nu\alpha} \,.
\end{equation}
Here, we have used the mimetic constraint (\ref{mimetic-const.}) to simplify the expression. The auxiliary field can be easily obtained from the trace of Einstein's equations as $\lambda=-G/4$, where $G=G^\mu_\mu$ is the trace of the Einstein tensor.


\subsection{Raychuadhuri equation}
Before restricting ourselves to a particular metric, we can understand singularity properties of the spacetime
 by considering the Raychuadhuri equation
\begin{eqnarray}\label{RaychuadhuriG}
\frac{d\theta}{d\tau}+\frac{1}{3}\theta^2=-\sigma_{\mu\nu}\sigma^{\mu\nu}
+\omega_{\mu\nu}\omega^{\mu\nu}-R_{\mu\nu}u^{\mu}u^{\nu}\,,
\end{eqnarray}
where $\sigma_{\mu\nu}$ is the shear tensor, $\omega_{\mu\nu}$ is the vorticity
tensor, $\theta = \nabla_{\mu} u^{\mu}$ is the expansion scalar and $u^{\mu}$ is the 
four-velocity that is a timelike vector field satisfying $u_{\mu}u^{\mu}=-1$. The expansion, shear, and vorticity would be obtained just after
choosing a particular metric. On the other hand, according to the Penrose-Hawking singularity theorem
\cite{Hawking:1973uf}, the singularity properties can be understood from the last term in 
(\ref{RaychuadhuriG}). As long as the strong energy condition is satisfied,  this term is always positive 
and therefore we conclude that the expansion rate diverges at some point, signalling the 
existence of a spacetime singularity. Therefore, we focus on this term to see whether it can 
be negative in our setup. From the Einstein equations we can write $R_{\mu\nu}u^{\mu}u^{\nu}
=(T_{\mu\nu}-\frac{1}{2}Tg_{\mu\nu})u^{\mu}u^{\nu}$ which, after substituting (\ref{EMT}), 
gives the following result
\begin{eqnarray}\label{Ruu0}
R_{\mu\nu}u^{\mu}u^{\nu} = - \frac{{\mathcal E}}{4} - 2 \lambda
+({\mathcal E}+4\lambda) \sum_{a}\, F_{\mu}^{(a)\alpha} F^{(a)}_{\nu\alpha} u^{\mu}u^{\nu}\,.
\end{eqnarray}

Employing the ADM decomposition of the metric
\begin{equation}
ds^2 = -N^2 dt^2 + q_{ij}(dx^i+N^idt)(dx^j+N^jdt)\,,
\end{equation}
the mimetic constraint (\ref{mimetic-const.}) can be decomposed into the electric and 
magnetic parts as follows
\begin{eqnarray}\label{mimetic-cons2}
-2 \sum_{a} q^{ij}F^{(a)}_{\perp i}F^{(a)}_{\perp j} + \sum_{a} q^{ik}q^{jl}F^{(a)}_{ij}F^{(a)}_{kl} = -1 \,,
\end{eqnarray}
where $F^{(a)}_{\perp i}\equiv n^{\mu}F^{(a)}_{\mu i}$ with $n^{\mu}\partial_{\mu}=(1/N)(\partial_t-N^i\partial_i)$, $F_{\perp}^{(a)i}\equiv q^{ij}F^{(a)}_{\perp j}$ and $q^{ij}$ is the inverse of $q_{ij}$. This relation allows us to rewrite the electric part $\sum_{a}q^{ij}F^{(a)}_{\perp i}F^{(a)}_{\perp j}$ in terms of the magnetic part $\sum_{a}q^{ik}q^{jl}F^{(a)}_{ij}F^{(a)}_{kl}$. In general $u^{\mu}$ is expanded as $u^{\mu}=-u^{\perp}n^{\mu}+\tilde{u}^i(\partial/\partial x^i)^{\mu}$ with $(u^{\perp})^2-q_{ij}\tilde{u}^i\tilde{u}^j=1$, where $u^{\perp}\equiv u^{\mu}n_{\mu}$ and $\tilde{u}^i\equiv q^{ij}u_j$. It is then straightforward to show that

\begin{eqnarray}\label{eqn:FFuu-general}
F_{\mu}^{(a)\alpha} F^{(a)}_{\nu\alpha} u^{\mu}u^{\nu}
&  = &
q^{ij}F^{(a)}_{\perp i}F^{(a)}_{\perp j} - 2u^{\perp}\tilde{u}^iq^{jk}F^{(a)}_{\perp j}F^{(a)}_{ik}
+ \tilde{u}^i\tilde{u}^jq^{kl}F^{(a)}_{ik}F^{(a)}_{jl} \nonumber\\
& & 
+ \tilde{u}^i\tilde{u}^j(q_{ij}q^{kl}F^{(a)}_{\perp k}F^{(a)}_{\perp l}
-F^{(a)}_{\perp i}F^{(a)}_{\perp j})\,.
\end{eqnarray}

For simplicity let us consider a vorticity-free flow with $u^{\mu}=n^{\mu}$. Using Eq. (\ref{mimetic-cons2}), Eq. (\ref{Ruu0}) is then rewritten as
\begin{eqnarray}\label{Ruu}
R_{\mu\nu}u^{\mu}u^{\nu}=\frac{{\mathcal E}}{4}\, + ({\mathcal E}+4\lambda)
\sum_{a}  q_{ij}B^{(a)i}B^{(a)j}\,,
\end{eqnarray}
where we have defined the magnetic field $B^{(a)i}\equiv \epsilon^{ijk}F^{(a)}_{jk}/2$ and $\epsilon^{ijk}$ is the $3$-dimensional Levi-Civita tensor with $\epsilon^{123}=1/\sqrt{\det q}$. This is interesting if we note that in 
the case of vanishing magnetic part $B^{(a)k}=0$ (or equivalently $F^{(a)}_{ij}=0$), the 
expression (\ref{Ruu}) reduces to
\begin{eqnarray}\label{Ruuc}
R_{\mu\nu}u^{\mu}u^{\nu}=\frac{\mathcal E}{4}\, = \mbox{constant} \,.
\end{eqnarray}

In the standard Einstein-Maxwell theory (with non-vanishing vorticity in general), both the electric part $F^{(a)}_{\perp i}$ and the magnetic part $F^{(a)}_{ij}$ contribute to $R_{\mu\nu}u^{\mu}u^{\nu}$. In our setup, the 
mimetic constraint (\ref{mimetic-cons2}) relates electric and magnetic parts to each other 
such that one can substitute one in terms of another. In general the second line of (\ref{eqn:FFuu-general}) remains but if it vanishes then (\ref{Ruuc}) holds in the case of vanishing magnetic part $B^{(a)k}=0$ (or equivalently $F^{(a)}_{ij}=0$).

The condition (\ref{Ruuc}) also holds for a maximally symmetric spacetime with constant 
curvature, if it admits a configuration for which the magnetic part of the field strength vanishes and the second line of (\ref{eqn:FFuu-general}) also vanishes.
For the case of curved spacetime with Lorentzian signature, in which we are 
interested here, there are two possibilities that are de Sitter and anti-de Sitter spacetimes. 
Regarding our setup with (\ref{Ruuc}), de Sitter and anti-de Sitter cases are corresponding to 
${\cal E}<0$ and ${\cal E}>0$ respectively. The de Sitter case ${\cal E}<0$ implies negative 
values for the last term in Raychuadhuri equation (\ref{RaychuadhuriG}) which shows that our 
model (\ref{action}) can be nonsingular in this case. The negative values ${\cal E}<0$ for the 
coupling constant of the Maxwell term in (\ref{action}) would not introduce any disastrous 
features. This can be easily deduced if we note that the Maxwell term in (\ref{action}) is 
forced to be constant through the mimetic constraint. 
Therefore, as we have explicitly shown in (\ref{Ruuc}), the coupling constant 
of the Maxwell term in mimetic gauge field model plays the role of the cosmological 
constant with the effective cosmological constant $\Lambda_{\rm eff}=-\frac{\cal E}{4}$.  For a de Sitter-like spacetime we require ${\cal E}<0$.

The above discussions show that demanding  the magnetic part of the gauge fields to vanish
$F^{(a)}_{ij}=0$, we can achieve a de Sitter-like solution in our model (\ref{action}) which can 
 also be nonsingular. We will see this fact in cosmological chart in the next section.


\subsection{Cosmological background equations}

In order to have a homogeneous and isotropic solution, we consider the following form 
for the components of the vector field \cite{Cervero:1978db,ArmendarizPicon:2004pm}
\begin{eqnarray}\label{vector-field}
A_{\mu}^{(a)}=A(t) \, \delta^a_\mu \,,
\end{eqnarray}
in which the function $A(t)$ completely determines the magnitudes of the vector fields. Although 
we have three different vector fields, we demand that their magnitudes to be 
exactly the same. In order to preserve this choice, which admits a homogeneous and isotropic solution, we have considered the global $O(3)$ symmetry in the field space.

The components of the field strength tensors (\ref{Fmunu-definition}) are given by
\begin{eqnarray}\label{Fmunu}
F^{(a)}_{0i} = \dot{A} \, \delta^a_j \,, \hspace{1cm} F^{(a)}_{ij}=0 \,.
\end{eqnarray}
Therefore, the magnetic part of the field strength tensors vanishes within the choice 
(\ref{vector-field}) for the vector fields. This is the desired condition to avoid the 
singularity in the presence of Maxwell term in our model (\ref{action}) which was already 
shown by means of the Raychuadhuri equation in (\ref{Ruu}).

We consider the spatially flat FRW spacetime with background metric
\begin{eqnarray}\label{FRW}
ds^2= - dt^2+ a(t)^2 \delta_{ij} dx^i dx^j \,,
\end{eqnarray}
where $a(t)$ denotes the scale factor. Substituting Eq. (\ref{Fmunu}) into the mimetic
constraints (\ref{mimetic-const.}) (or equivalently in Eq. (\ref{mimetic-cons2})), we find that 
\begin{eqnarray}\label{Adot}
\dot{A}=\frac{a(t)}{\sqrt{6}} \,,
\end{eqnarray}
where we have assumed that $\dot{A}\geq 0$ without loss of generality.

We now consider the Einstein's equations in the cosmological background (\ref{FRW}) under the condition (\ref{Adot}).
The $00$ component gives the Friedmann equation
\begin{eqnarray}\label{Friedmann0}
3 H^2 = 2 \lambda + \frac{\mathcal{E}}{4}\,,
\end{eqnarray}
where $H(t)=\dot{a}(t)/a(t)$ denotes the Hubble parameter.

The $ii$ components give the dynamical equation
\begin{eqnarray}\label{Raychuadhuri}
2 \dot{H} + 3 H^2 = \frac{2}{3} \lambda - \frac{\mathcal{E}}{12} \,.
\end{eqnarray}
From the above equations, we find the following solution for the auxiliary field $\lambda$
\begin{eqnarray}\label{sollambda}
\lambda = - \frac{\mathcal{E}}{4} - \frac{3}{2} \dot{H} \,.
\end{eqnarray}

In the following we consider two cases of ${\mathcal{E}}=0$ and ${\mathcal{E}} \neq 0$. The
first is the pure mimetic effect while the latter includes the effect of the Maxwell term which would
mimic the roles of cosmological constant in our model through the mimetic constraint. 


\subsubsection{${\mathcal{E}}=0$: Mimetic energy density as a spatial curvature}

In the case of ${\mathcal{E}}=0$, from Eqs. (\ref{Friedmann0}) and (\ref{Raychuadhuri}), we
easily find that the setup describes a perfect fluid with energy density $\rho=2\lambda$ and 
equation of state parameter $w=-\frac{1}{3}$ which decays as
\begin{eqnarray}\label{curvature-energydensity}
\rho \sim {a^{-2}} \,.
\end{eqnarray}
Thus the mimetic $1$-form model behaves as an effective spatial curvature. This is in a sharp contrast to the mimetic $0$-form model (\ref{action-p0}), i.e. the original mimetic model \cite{Mimetic-2013}, that reproduces a dark matter-like energy density component $\rho\sim{a^{-3}}$ \cite{Mimetic-2013}.

Therefore, although we have started from a spatially flat FRW universe (\ref{FRW}), the mimetic constraint, the second term in (\ref{action}), mimics the roles of an effective spatial curvature.


\subsubsection{${\mathcal{E}} \neq 0$: de Sitter universes}

The Maxwell term in (\ref{action}) is forced to be constant through the mimetic constraint 
(\ref{mimetic-const.}) and therefore it is natural to expect that ${\mathcal E}$ plays the roles of 
cosmological constant as we have already shown in Eq. (\ref{Ruuc}). Indeed, through a redefinition of the Lagrange 
multiplier $\lambda$, the Maxwell term can be replaced by a constant term at the level of the action. In order to explicitly see this fact at the level of the equations of motion, we substitute (\ref{sollambda}) into the Friedmann equation (\ref{Friedmann0}) which gives 
\begin{eqnarray}\label{BGE-pre}
\dot{H}+H^2= \frac{\Lambda_{\rm eff}}{3} \,,
\end{eqnarray}
where the effective cosmological constant is defined as
\begin{eqnarray}\label{effectiveCC}
\Lambda_{\rm eff}= - \frac{\cal E}{4} \,.
\end{eqnarray}
By integrating Eq. (\ref{BGE-pre}) once, we obtain
\begin{equation} \label{BGE}
3\left(H^2+\frac{K_{\rm eff}}{a^2}\right) = \Lambda_{\rm eff}\,,
\end{equation}
where $K_{\rm eff}$ is an integration constant.

We have started with a spatially flat FRW universe in Eq. (\ref{FRW}), but Eq. (\ref{BGE}) is the same for the  flat, closed and open de Sitter universes with $K_{\rm eff}=0$, $>0$ and $<0$, respectively. This is not surprising if we note that, as demonstrated in the previous subsection,  the mimetic constraint term behaves exactly the same as the spatial curvature. The integration constant $K_{\rm eff}$ plays the role of the effective spatial curvature and determines the sign of $\dot{H}$ such that $K_{\rm eff}=0$, $<0$ and $>0$ correspond respectively to flat, open and closed de Sitter-like universes which we study in turns below.

\paragraph{Flat de Sitter universe:}

From equation (\ref{sollambda}), we can see that the flat de Sitter solution 
$K_{\rm eff}=0$ corresponds to $\lambda = - \frac{\cal E}{4}$. Solving Eq. (\ref{BGE}) in this
case, we obtain the well-known results 
\begin{eqnarray}
a(t) & = & \exp(H_{\Lambda}t)\,, \label{scalefactor-flat} \\
H(t) & = & H_{\Lambda}\,, \label{Hubble-flat}
\end{eqnarray}
where
\begin{eqnarray}\label{Hmax}
H_{\Lambda}^2=\frac{\Lambda_{\rm eff}}{3} \,,
\end{eqnarray} 
We have chosen the origin of the time coordinate $t$ such that $a(0)=1$. This solution is maximally symmetric and is nothing but the de Sitter spacetime. 

\paragraph{Open de Sitter-like universe:}

Solving Eq. (\ref{BGE}) with $K_{\rm eff}=-1$, we obtain
\begin{eqnarray}
a(t) & = & \frac{1}{H_{\Lambda}} \sinh(H_{\Lambda}t)\,, \label{scalefactor-open} \\
H(t) & = & H_{\Lambda} \coth(H_{\Lambda}t)\,, \label{Hubble-open}
\end{eqnarray}
where $H_{\Lambda}$ is defined in (\ref{Hmax}) and, similar to the flat de Sitter-like case, 
we have chosen the origin of the time coordinate $t$ such that $a(0)=0$. 

The time derivative of the Hubble expansion rate then turns out to be
\begin{eqnarray}\label{Hubbledot-open}
\dot{H}(t) = - H_{\Lambda}^2  \csch^2(H_{\Lambda}t) <0 \,.
\end{eqnarray}
The Hubble expansion rate is bounded from below but unbounded from above as $H_{\Lambda}\leq H <\infty$. At the time $t=0$, the universe hits the singularity such that the scale factor Eq. (\ref{scalefactor-open}) vanishes and the Hubble expansion rate Eq. (\ref{Hubble-open}) diverges. This solution is not maximally symmetric since $K_{\rm eff}$ does not stem from the spatial curvature.

\paragraph{Closed de Sitter-like universe:}

Now, we focus on the case with $K_{\rm eff}>0$ in which the singularity can be avoided. Solving (\ref{BGE}) with $K_{\rm eff}=1$, we obtain
\begin{eqnarray}
 a(t) & = & \frac{1}{H_{\Lambda}} \cosh(H_{\Lambda}t)\,,\label{scalefactor-closed}\\
 H(t) & = & H_{\Lambda} \tanh(H_{\Lambda}t)\,,\label{Hubble-closed}
\end{eqnarray}
in which $H_{\Lambda}$ is given by Eq. (\ref{Hmax}) and we have chosen the origin of the time such that $H(0)=0$. Again, this solution is not maximally symmetric since $K_{\rm eff}$ does not stem from the spatial curvature. 

In order to study the bouncing behavior of the model, it is useful to look at the time 
derivative of the Hubble expansion rate which is given by
\begin{eqnarray}\label{Hubbledot-closed}
\dot{H}(t) = H_{\Lambda}^2  \sech^2(H_{\Lambda}t) = H_{\Lambda}^2 - H^2 > 0 \,.
\end{eqnarray}
From Eq. (\ref{Hubble-closed}), it is clear that the Hubble expansion rate is bounded as
\begin{eqnarray}\label{HUbble-Boundedness}
-H_{\Lambda}<H<H_{\Lambda}\,,
\end{eqnarray}
and it approaches $\pm{H_{\Lambda}}$ as $t\rightarrow\pm\infty$, where Eq. (\ref{Hubbledot-closed}) shows that $\dot{H}\rightarrow0$. In this limit the universe undergoes an exponential expansion with constant Hubble expansion rate. On the other hand, the Hubble expansion rate vanishes at $t=0$ where the scale factor (\ref{scalefactor-closed}) approaches its nonzero minimum value $a_{\rm min}=H_{\Lambda}^{-1}$. The universe starts from the past infinity $t\rightarrow{-\infty}$ with $H=-H_{\Lambda}<0$ and then bounces from the contracting phase with $-H_{\Lambda}<H<0$ into the expanding era with $0<H<H_{\Lambda}$ at $t=0$. Finally, at the future infinity $t\rightarrow+\infty$, the Hubble expansion rate approaches the constant maximum value $H=H_{\Lambda}$ with $\dot{H}=0$ and a de Sitter phase with effective cosmological constant (\ref{effectiveCC}) and exponentially growing scale factor $a(t)=e^{H_{\Lambda}t}$ arises in this scenario.

From Eq. (\ref{Hubbledot-closed}), it is clear that the Hubble expansion rate always grows,  $\dot{H}>0$. On the other hand, $\dot{H}=-\frac{1}{2}(\rho+p)$ and therefore we have $\rho+p<0$ so the null energy condition (NEC) is violated.  Indeed, the violation of NEC is the common feature of bouncing models \cite{Khoury:2001bz,Battefeld:2014uga} and it is commonly associated with ghost or gradient instabilities in scalar modes at the linear perturbations level \cite{Cline:2003gs}. It is also possible to construct some bouncing models which violate the NEC without introducing pathologies \cite{Creminelli:2006xe}. In the next section, we will see that  the scalar and vector modes suffer from ghost instabilities
whenever $\dot{H}>0$.

\subsection{Cosmological perturbations}
\label{cosmological-perturbations}

In order to study the stability of our setup with the action Eq. (\ref{action}), in this section we perform its linear perturbations analysis.

The metric perturbations around the background geometry Eq. (\ref{FRW}) are given by
\begin{eqnarray}\label{metric-perturbations}
\delta{g_{00}} = 2 \alpha \,, \hspace{.5cm} \delta{g_{0i}} = a^2 (\partial_i \beta + B_i )\,,
\hspace{.5cm} \delta{g_{ij}} = a^2 (2\psi \delta_{ij} + 2\partial_i\partial_j E 
+ \partial_i F_j + \partial_j F_i +h_{ij}) \,,
\end{eqnarray}
while the perturbations in gauge field with the global $O(3)$ symmetry are given by 
\cite{Emami:2016ldl}
\begin{eqnarray}\label{GF-perturbations}
\delta{A_0^{(a)}} = Y_a+\partial_a Y \,, \hspace{.5cm}
\delta{A_i^{(a)}} = a(t) \big[ \delta{Q}\, \delta_{ia}+\partial_i (\partial_a M+M_a)
+ \epsilon_{iab}(\partial_b{U}+U_b) + t_{ia} \big] \,,
\end{eqnarray}
where $(\alpha,\beta,\psi,E,Y,\delta{Q},M,U)$ are scalar modes, $(B_i,F_i,Y_a,M_a,U_a)$ are vector modes, and $(h_{ij},t_{ia})$ label tensor modes. The vector and tensor modes satisfy the transverse and traceless conditions
\begin{eqnarray}\label{transverse-traceless}
\partial_i B_i = \partial_i F_i = \partial_i Y_i = \partial_i M_i = \partial_i U_i = 0 \,,  \hspace{.5cm}
\partial_i h_{ij} = \partial_i t_{ij} = 0 \,, \hspace{.5cm} h_{ii} = t_{ii} = 0 \,.
\end{eqnarray}

The diffeomorphism invariance associated with the general coordinate transformation fixes two scalar modes and 
two vector modes. For the scalar modes, we work in the spatially flat gauge with $\psi=E=0$ and 
for the case of vector modes we fix the gauge as $F_i=0$. Moreover, the model (\ref{action}) 
enjoys the local gauge symmetry 
$A^{(a)}_{\mu}\rightarrow{A}^{(a)}_{\mu}-\partial_{\mu}\Lambda^a$. Decomposing $\Lambda^a $ into $\Lambda^a 
= \partial_a \Lambda + \Lambda^{\perp}_a$ with $\partial_a\Lambda^{\perp}_a=0$, at the level 
of linear perturbations we have
\begin{eqnarray}\label{LGS}
\delta{A}^{(a)}_{\mu}\rightarrow\delta{A}^{(a)}_{\mu} - 
\partial_{\mu}\partial_a \Lambda-\partial_{\mu}\Lambda^{\perp}_a \,,
\end{eqnarray}
which for perturbations in Eq. (\ref{GF-perturbations}) implies
\begin{eqnarray}\label{LGF-perturbations}
&& Y \rightarrow Y - \dot{\Lambda}\,, \hspace{.5cm} M \rightarrow M - a^{-1} \Lambda \,, \\
&& Y_a \rightarrow Y_a - \dot{\Lambda}^{\perp}_a\,, \hspace{.5cm} 
M_a \rightarrow M_a - a^{-1} \Lambda^{\perp}_a \,. \nonumber
\end{eqnarray}
All the other perturbations in Eq. (\ref{GF-perturbations}) are invariant under the local gauge transformation (\ref{LGS}). The above relations show that one scalar mode and two vector modes are not real physical degrees of freedom and can be removed through the local transformation (\ref{LGS}). Thus, without loss of generality, we choose $M=0$ and $M_a=0$. 

Apart from the above gauge degrees of freedom, the perturbations are also restricted by the
mimetic constraint (\ref{mimetic-const.}), which at the linear order implies  
\begin{eqnarray}\label{mimetic-const.-LP0}
\sqrt{\frac{3}{2}} a \, \alpha + \partial^2 Y - 3 \delta{\dot Q} - \partial^2 {\dot M}
+ \partial_i Y_i - \partial_i {\dot M}_i + \frac{a}{2\sqrt{6}} h_{ii} - t_{ii} = 0 \,.
\end{eqnarray}

Imposing the transverse and traceless conditions (\ref{transverse-traceless}), we can easily 
see that the mimetic constraint induced by (\ref{mimetic-const.-LP0}) gives a relation among scalar modes in which 
 we can write one scalar mode in terms of others as 
 \begin{eqnarray}\label{mimetic-const.-LP1}
\alpha = \sqrt{\frac{2}{3}} a^{-1} \left(- \partial^2 Y + 3 \delta{\dot Q} 
+ \partial^2 {\dot M} \right)\, .
\end{eqnarray}
Setting $M=0$ by the local gauge transformation Eq. (\ref{LGF-perturbations}), this yields 
\begin{eqnarray}\label{mimetic-const.-LP}
\alpha = \sqrt{\frac{2}{3}} a^{-1} \left(- \partial^2 Y + 3 \delta{\dot Q} \right)\,.
\end{eqnarray}

In conclusion, after fixing all gauge freedoms and imposing the mimetic constraint we are left with four
scalar modes $(\beta,Y,\delta{Q},U)$, six vector modes $(B_i,Y_a,U_a)$, and four tensor
modes $(h_{ij},t_{ia})$. It turns out that the scalar modes $(\beta,Y)$ and the vector modes 
$(B_i,Y_a)$ are non-dynamical. Therefore, the quadratic action takes the following form
\begin{eqnarray}\label{actionO20}
S^{(2)} =  S^{(2)}_S\big(\delta{Q},U\big) + S^{(2)}_V\big(U_a\big) + S^{(2)}_T\big(h_{ij},t_{ij}\big) \,.
\end{eqnarray}
Note that the scalar, vector, and tensor modes are decoupled from each other, thanks to the rotational symmetry of the background and the global $O(3)$ symmetry in the field space of the model given by the action (\ref{action}). 

Below we study each type of perturbations separately. 

\subsubsection{Scalar perturbations}

Going to the Fourier space, after some calculations, it is straightforward to show that 
the quadratic action for the scalar modes is given by\footnote{For the sake of simplicity, we denote the Fourier amplitude $A_{k}(t)$ by $A$ for all of the perturbations.}
\begin{eqnarray}\label{actionO2S}
S^{(2)}_S\big(\delta{Q},U\big) = 6 \int d^3k dt \, a^3 \left( - \dot{H}\right) 
\Big[ 3 \Big(\dot{\delta{Q}}^2 - \frac{k^2}{3 a^2} \delta{Q}^2
+ \frac{\dot H}{H^2} (2\dot{H}+3H^2)\delta{Q}^2 \Big) 
\\ \nonumber
+ k^2 \Big( \dot{U}^2 - \frac{k^2}{a^2} U^2 - \dot{H} U^2 \Big) \Big] \,.
\end{eqnarray}

From the above action, we can easily see that both scalar modes $(\delta{Q},U)$ become
ghost if we consider the bouncing (closed de Sitter-like) solution Eq. (\ref{Hubbledot-closed}) with
$\dot{H}>0$. On the other hand, both of them are free of any ghost and gradient/Laplacian 
instabilities for $\dot{H}<0$. The mode $U$ may exhibit a tachyonic instability in infrared (IR) regime $k\rightarrow0$ even for $\dot{H}<0$. Such infrared instabilities are similar to the standard Jeans gravitational instabilities and are not necessarily a real pathology of the model \cite{Gumrukcuoglu:2016jbh}. Moreover, in the ultraviolet (UV)  regime, the scalar mode $\delta{Q}$ propagates with the squared sound speed  $c_{s}^2=\frac{1}{3}$ while $U$ propagates with the speed of light. Finally, for $\dot{H}=0$, the coefficients of the kinetic terms for the two scalar modes vanish and thus they may signal a strong coupling, depending on the behavior of nonlinear interactions. 

\subsubsection{Vector perturbations}

With the same manner as in scalar perturbations, we can find the quadratic action for the vector 
modes in the Fourier space as 
\begin{eqnarray}\label{actionO2V}
S^{(2)}_V\big(U_a\big) =  9 \sum_{a=1,2} \int d^3k dt \, a^3 \left( - \dot{H}\right) 
\Big[ \dot{U}_a^2 - \frac{k^2}{3 a^2} U_a^2 - \frac{5}{3} H^2 U_a^2 \Big] \,.
\end{eqnarray}

The two vector modes become ghost for the bouncing (closed-de Sitter like) solution (\ref{Hubbledot-closed}) with $\dot{H}>0$ while they are free of any ghost and gradient/Laplacian instabilities for $\dot{H}<0$. They also propagate with the squared sound speed $c_{v}^2=\frac{1}{3}$ in the UV regime and they have mass proportional to the Hubble expansion rate in the IR regime. 

\subsubsection{Tensor perturbations}

The tensor sector of the model consists of $h_{ij}$ from the metric and $t_{ij}$ from the gauge fields and they couple with each other. It is straightforward to show that the quadratic action for the tensor modes in the Fourier space is 
\begin{eqnarray}\label{actionO2T}
S^{(2)}_T\big(h_{ij},t_{ij}\big) = \frac{1}{2} \int d^3k dt\, a^3 \Big( \dot{\bf X}^T {\bf K} \dot{\bf X}
+  \dot{\bf X}^T {\bf N} {\bf X} - {\bf X}^T {\bf M} {\bf X} \Big) \,,
\end{eqnarray}
where ${\bf X}$ is a column vector with the component values $X^i=(h_{11},h_{12},t_{11},t_{12})^T$ 
and ${\bf X}^T$ denotes its transpose. The $4\times4$ symmetric kinetic matrix ${\bf K}$ 
determines the coefficients of all kinetic terms which has nonzero components
\begin{eqnarray}\label{K}
K_{11}=K_{22}=1 \,, \hspace{.5cm} K_{33} = K_{44}= -24 \dot{H} \,, 
\end{eqnarray}
and the $4\times4$ matrix ${\bf N}$ has nonzero components 
\begin{eqnarray}\label{N}
N_{13}=N_{24}= - 4 \sqrt{6} \dot{H}\, .
\end{eqnarray}
The coefficients of the gradient and mass terms are determined by $4\times4$ symmetric diagonal 
matrix ${\bf M}$ with the following nonzero components 
\begin{eqnarray}\label{M}
M_{11}=M_{22}=\frac{k^2}{a^2} + 2 \dot{H} \,, \hspace{.5cm} 
M_{33} = M_{44} = -24 \dot{H} \Big(\frac{k^2}{a^2} + \dot{H} \Big) \,. 
\end{eqnarray}

Regarding the ghost instabilities, all of the information are encoded in the kinetic matrix ${\bf K}$ which is diagonal. Therefore each diagonal element in Eq. (\ref{K}) determines whether the corresponding mode is plagued with ghost-like instability or not. Since all of the elements are positive in the case of $\dot{H}<0$, there would be no ghost instabilities for the tensor modes for $\dot{H}<0$. However, for the bouncing (closed-de Sitter like) solution (\ref{Hubbledot-closed}) with $\dot{H}>0$, two of tensor modes associated with the gauge field $t_{ij}$ become ghost.

The computation of the  sound speeds is also easy. In the subhorizon limit, $k^2/a^2\gg H^2$, all components of the friction matrix ${\bf N}$ are $\mathcal{O}(k^0)$ and thus do not contribute to the squared sound speeds. In the same limit the mass matrix ${\bf M}$ is 
\begin{eqnarray}
 {\bf M} = \frac{k^2}{a^2} {\bf K}  + \mathcal{O}(k^0)\,.
\end{eqnarray}
Therefore, all four tensor modes propagate with the speed of light in the subhorizon limit and there is no Laplacian instability in the tensor sector.


\section{Summary and Discussions}
\label{summary}

In the original mimetic dark matter scenario, the conformal degree of freedom of the metric is isolated by means of a scalar field. In this work, we have extended the original 
mimetic scenario to general gauge invariant $p$-form scenarios. The $0$-form case 
with scalar potential reproduces the original mimetic dark matter scenario. The 
$1$-form case corresponds to a vector potential in which the conformal degree of 
freedom is isolated by the strength tensor of the vector potential. We have explicitly 
confirmed this fact by looking at the singular limit of the associated disformal 
transformation. We have shown that the remaining $2$-form mimetic model is 
equivalent to the standard mimetic $0$-form model through the Hodge duality.

We then studied the cosmological implications of the $1$-form model. Considering a 
global $O(3)$ symmetry for the gauge field to allow for isotropic FRW backgrounds, we 
have obtained the associated cosmological background solutions. In comparison with 
the standard mimetic $0$-form model where the scalar field produces a dark matter-like 
energy density component, we have found that the $1$-form model produces energy 
density component like the spatial curvature. Due to the mimetic constraint, the 
Maxwell term (and any function of it) behaves like the cosmological constant term. 
Adding the Maxwell term we then found, flat, open and closed de Sitter-like solutions. Moreover, for the closed de Sitter-like setup we can obtain bouncing solution.  
Performing perturbations analysis, however, we have shown that the closed de Sitter-like solution suffers from ghost instabilities while the open de Sitter-like solution 
is stable. In the case of flat de Sitter-like solution, we found that the quadratic actions
for the scalar and vector modes together with gauge field tensor modes vanish which,
depending on the behavior of nonlinear interactions, may signal that the model is 
strongly coupled.

While our analysis show that the setup with $\dot{H}>0$ suffers from ghost 
instabilities, one may wonder whether the Horndeski's non-minimal coupling term \cite{Horndeski:1974wa} 
\begin{eqnarray}\label{Horndeski-NC}
\sum_{a=1}^3\left(\,R F^{(a)}_{\mu\nu} F^{(a)\mu\nu} \,-\, 4 R_{\mu\nu} F^{(a)\mu}_{\alpha} F^{(a)\nu\alpha} \,+\, R_{\mu\nu\alpha\beta} F^{(a)\mu\nu} F^{(a)\alpha\beta} \, \right) \,,
\end{eqnarray}
can remove the instabilities and provide stable bouncing solution. We have considered
the effects of the above term on the stability analysis of our model, and have found that it is not 
possible to construct a stable bouncing model in our scenario. Specifically, our results 
show that we can find a region in the  parameter space in which the scalar and vector modes 
are free of any ghost and gradient/Laplacian instabilities but the tensor sector is always sick in the bouncing background. With the mimetic costraint (\ref{mimetic-const.}), we could generalize the Horndeski's non-minimal coupling (\ref{Horndeski-NC}) by allowing the coefficient of the first term to be arbitrary but this generalization is equivalent to a redefinition of the Lagrange multiplier $\lambda_1$ and thus does not change the conclusion.

As we have seen, the  effective energy density from the gauge field mimetic constraint mimics the roles of 
the spatial curvature in our setup. This inspires us to extend the setup to the spatially curved FRW universe. 
In order to do this, a simple analysis shows that we need to consider the $SU(2)$ gauge symmetry instead of the $U(1)\times U(1)\times U(1)$ gauge symmetry. In the limit of zero
gauge coupling constant, the model would reduce to the current global $O(3)$ model and therefore 
we expect to recover all of our results here. We would like to come back to this question in the near future.   


\vspace{0.7cm}

{\bf Acknowledgments:}  We would like to thank Javad T. Firouzjaee for useful discussions. MAG and HF would like to thank the Yukawa Institute for Theoretical Physics at Kyoto University for hospitality.  Discussions during the YITP symposium YKIS2018a ``General Relativity -- The Next Generation --" were useful for the  completion of this work. The work of SM was supported by Japan Society for the Promotion of Science (JSPS)  Grants-in-Aid for Scientific Research (KAKENHI) No. 17H02890, No. 17H06359, and by 
World Premier International Research Center Initiative (WPI), MEXT, Japan.  


\vspace{0.7cm}

\appendix

\section{Singular limit of gauge-invariant conformal transformation}
\setcounter{equation}{0}
\renewcommand{\theequation}{A\arabic{equation}}

In the Introduction Section, we have mentioned  that Eq. (\ref{mimetic-const.0}) would be 
the natural gauge field extension of the scalar mimetic constraint Eq. (\ref{mimetic-scalar-cons.}). Here we
prove our claim through the same line that Eq. (\ref{mimetic-scalar-cons.}) has been realized
as the singular limit of a conformal/disformal transformation \cite{Deruelle:2014zza}.

In order to do this, we consider the following conformal transformation between the physical
metric $g_{\mu\nu}$ and auxiliary metric ${\tilde g}_{\mu\nu}$
\begin{eqnarray}\label{conformal-tarns}
g_{\mu\nu} = \, A(X) \, {\tilde g}_{\mu\nu} \,,
\end{eqnarray} 
where
\begin{eqnarray}
X = -\tilde{g}^{\mu\alpha} \tilde{g}^{\nu\beta} F_{\alpha\beta} F_{\mu\nu} \,.
\end{eqnarray} 
Note that the conformal transformation (\ref{conformal-tarns}) is gauge-invariant. Our task is 
to see whether the transformation Eq. (\ref{conformal-tarns}) is invertible, {\it i.e.} if we can find the
auxiliary metric in terms of the physical metric ${\tilde g}_{\mu\nu}(g_{\mu\nu})$. Since the
coefficient $A$ in Eq. (\ref{conformal-tarns}) is a function of the auxiliary metric, we should look 
at the Jacobian of the transformation. Equivalently, we can study the following eigenvalue  
problem for the determinant of the Jacobian \cite{Zumalacarregui:2013pma}
\begin{eqnarray}\label{eigenvalue-eq}
\left( \frac{\partial{g_{\mu\nu}}} {\partial {\tilde g}_{\alpha\beta}} 
- \eta_i\, \delta_{\mu}^{\alpha} \delta_{\nu}^{\beta} \right) \xi^i_{\alpha\beta} = 0 \,,
\end{eqnarray}
in which $\eta_i$ are the eigenvalues and $\xi^i_{\alpha\beta}$ are the associated eigentensors. 

From Eq. (\ref{conformal-tarns}), we can easily find that
\begin{eqnarray}\label{Jacobian}
\frac{\partial{g_{\mu\nu}}} {\partial {\tilde g}_{\alpha\beta}} = 
A\, \delta_{\mu}^{\alpha} \delta_{\nu}^{\beta} 
- 2 A_{,X} F^{\alpha\lambda} F_\lambda^{\,\,\beta} {\tilde g}_{\mu\nu} \,,
\end{eqnarray}
where the indices in the r.h.s. are raised and lowered by auxiliary metric 
${\tilde g}_{\mu\nu}$. Substituting the above result in Eq. (\ref{eigenvalue-eq}) gives
\begin{eqnarray}\label{eigenvalue-eq2}
(A-\eta_i) \xi^i_{\mu\nu} 
- 2 A_{,X} F^{\alpha\lambda} F_\lambda^{\,\,\beta} \xi^i_{\alpha\beta} 
{\tilde g}_{\mu\nu} = 0 \,.
\end{eqnarray}
There are two sets of solutions for the above eigenvalue problem
\begin{eqnarray}\label{eigenvalues}
&& \eta_C = A \,; \hspace{2.8cm} \xi^C_{\mu\nu} = w_{\mu\nu} \,, 
\hspace{.3cm}\mbox{with}\hspace{.3cm} 
F^{\alpha\lambda} F_\lambda^{\,\,\beta} w_{\alpha\beta} = 0 \,, \nonumber \\
&& \eta_X = A - 2 X A_{,X} \,; \hspace{1cm} \xi^X_{\mu\nu} = {\tilde g}_{\mu\nu} \,,
\end{eqnarray}
where, following the convention of  \cite{Firouzjahi:2018xob}, the  superscripts $C$ and $X$ denote the 
conformal-type and the kinetic-type eigenvalues. Note that the 
conformal-type eigenvalue is degenerate with multiplicity of nine since there is one constraint while
the kinetic-type eigenvalue is non-degenerate. The singular limit of the transformation 
Eq. (\ref{conformal-tarns}) is given by $\eta_i=0$. In the case of conformal-type eigenvalue $\eta_C=0$, 
we obtain the trivial solution $A=0$  while for the case of kinetic-type eigenvalue $\eta_X=0$  we find the following nontrivial solution
\begin{eqnarray}
A = \sqrt{X} \,,
\end{eqnarray} 
where, without loss of generality, we have set the constant of integration to be unity.
Substituting the above result in the conformal transformation Eq. (\ref{conformal-tarns}), we 
obtain Eq. (\ref{mimetic-trans-GF}). 

Therefore, the transformation Eq. (\ref{mimetic-trans-GF}) is the natural extension of the scalar 
mimetic transformation Eq. (\ref{mimetic-trans}) to the case of gauge field. We have used this 
analogy in Section (\ref{pform}) in order to investigate the mimetic constraint 
Eq. (\ref{mimetic-const-p}) in terms of a general $p$-form.

\section{The equivalence of $p=0$ and $p=2$ models}

\setcounter{equation}{0}
\renewcommand{\theequation}{B\arabic{equation}}

In this appendix, we show that the mimetic $0$-form and $2$-form models defined by the 
actions (\ref{action-p0}) and (\ref{action-p2}) are equivalent to each other. More precisely, 
using the Hodge duality, we show that the field strengths of the two models satisfy the 
same equations of motion.

The $1$-form field strength for the $p=0$ model with $0$-form scalar potential $\phi$ is 
given by $\Phi=d\phi$ and the Bianchi identity $d\Phi=0$ implies
\begin{eqnarray}\label{0form-Eq1-Bianchi}
\nabla_{\mu}\Phi_{\nu} - \nabla_{\nu}\Phi_{\mu} = 0 \,,
\end{eqnarray}
where $\Phi_{\mu}=\partial_{\mu}\phi$ are the components of the field strength tensor. 
The variation of the action (\ref{action-p0}) with respect to $\phi$ gives the following modified 
Klein-Gordon equation
\begin{eqnarray}\label{0form-Eq2-KG}
\nabla_{\alpha} \left(\lambda_0 \Phi^{\alpha}\right) = 0 \,.
\end{eqnarray}
Moreover, the variation with respect to the metric gives the 
energy-momentum tensor
\begin{eqnarray}\label{0form-Eq3-EMT}
T^{\mu}_{\nu} = 2\lambda_0 \Phi^{\mu} \Phi_{\nu} \,,
\end{eqnarray}
in which we have used the mimetic constraint Eq. (\ref{mimetic-const-p}) in the case of $p=0$. 

In the case of $p=2$ with $2$-form potential 
$B=\frac{1}{2!}B_{\mu\nu}dx^{\mu}\wedge dx^{\nu}$, the associated $3$-form field strength 
is given by $J=dB$ and the Bianchi identity $dJ=0$ then implies that 
\begin{eqnarray}\label{2form-Eq1-Bianchi}
\nabla_{\alpha} J_{\beta\mu\nu} - \nabla_{\beta} J_{\mu\nu\alpha} 
+ \nabla_{\mu} J_{\nu\alpha\beta} - \nabla_{\nu} J_{\alpha\beta\mu} = 0 \,.
\end{eqnarray}
Varying the action (\ref{action-p2}) with respect to the $2$-form potential $B$, we obtain 
\begin{eqnarray}\label{2form-Eq2-KG}
\nabla_{\alpha} \left(\lambda_2 J^{\alpha\mu\nu}\right) = 0 \, ,
\end{eqnarray}
while the variation with respect to the metric gives the 
corresponding energy-momentum tensor as
\begin{eqnarray}\label{2form-Eq3-EMT}
T^{\mu}_{\nu} = \lambda_2 J^{\mu\alpha\beta} J_{\nu\alpha\beta} \,,
\end{eqnarray}
where $J_{\mu\alpha\beta}$ are the components of the field strength tensor.

Rewriting Eqs. (\ref{2form-Eq1-Bianchi}), (\ref{2form-Eq2-KG}), and (\ref{2form-Eq3-EMT}) in
terms of the dual of the strength tensor $J$ which is defined as 
$J_{\nu}=\frac{1}{3!}\epsilon_{\alpha\beta\mu\nu} J^{\alpha\beta\mu}$, we have
\begin{eqnarray}\label{2form-Eq1-dual}
\nabla_{\alpha} J^{\alpha} = 0 \,,
\end{eqnarray}
\begin{eqnarray}\label{2form-Eq2-dual}
\nabla_{\mu} \left(\lambda_2 J_{\nu}\right) - \nabla_{\nu} \left(\lambda_2 J_{\mu}\right) = 0 \,,
\end{eqnarray}
\begin{eqnarray}\label{2form-Eq3-dual}
T^{\mu}_{\nu} = 2 \lambda_2 J^{\mu} J_{\nu} \,.
\end{eqnarray}
Comparing (\ref{2form-Eq2-dual}) with (\ref{0form-Eq1-Bianchi}), we see that the two 
equations are the same if we consider the identification
\begin{eqnarray}\label{duality-Eq1}
\Phi_{\mu}  \,\leftrightarrow\, \lambda_2 J_{\mu} \,.
\end{eqnarray}
Substituting the above relation into the equations (\ref{2form-Eq1-dual}) and 
(\ref{2form-Eq3-dual}) and then comparing the results with Eq. (\ref{0form-Eq2-KG}) and 
Eq. (\ref{0form-Eq3-EMT}), we find the following correspondence between the Lagrange multipliers
\begin{eqnarray}\label{duality-Eq2}
\lambda_0 \,\leftrightarrow\, \frac{1}{\lambda_2} \,.
\end{eqnarray}
In some sense, the above relation shows the weak and strong duality between the $0$-form and
$2$-form models. 

In conclusion, we have shown that  the $2$-form model is the Hodge dual of the
standard mimetic scenario which is given by the $0$-form model in our classification. 

{}

\end{document}